\documentstyle[12pt,psfig]{article}
\begin{document}
\vspace*{1.5cm}
\begin{center} 
\Large{\bf Energy Levels of Interacting Fields in a Box} 
\\ [10mm] 
\end{center}
\renewcommand{\thefootnote}{\alph{footnote}} 
\begin{center}
{\bf J. A. Espich\'an Carrillo,$^{1}$ 
and  
A. Maia Jr.$^{2}$} 
\end{center} 




\vspace*{0.5cm}

\begin{center}
\parbox{15cm}{We study the influence of boundary conditions on energy levels of interacting fields in a box and discuss some consequences when we change the size of the box. In order to do this we calculate the energy levels of bound states of a scalar massive field $\chi$ interacting with another scalar field $\phi$ through the lagrangian ${\cal L}_{int} = \frac{3}{2}g\phi^{2}\chi^{2}$ in an one-dimensional box, on which we impose Dirichlet boundary conditions.
We have found that the gap between the bound states changes with 
the size of the box in a non-trivial way. For the case the masses of the two fields are equal and for large box the energy levels of 
Dashen-Hasslacher-Neveu (DHN model) (Dashen {\it et al}, 1974) are recovered and we have a kind of boson condensate for the ground state. Below to a critical box size $L\sim 2.93\,\frac{2\sqrt{2}}{M}$ the ground state level splits, which we interpret as particle-antiparticle production under small perturbations of box size. Below another critical sizes $(L\sim \frac{6}{10}\frac{2\sqrt{2}}{M})$ and $(L\sim 1.71\frac{2\sqrt{2}}{M})$ of the box, the ground state and first excited state merge in the continuum part of the spectrum.}
\end{center}

\vspace*{0.3cm}

\noindent
key words: Energy Levels,  Particle Production, Interacting Fields, 
Quantum Field Theory.


\baselineskip 0.65cm


\vspace*{1cm}

\noindent
{\large \bf 1. INTRODUCTION}

\vspace*{0.5cm}

Consider a simple system of two interacting fields, described by the lagrangian density
\begin{eqnarray*}
{\cal L} =  \frac{1}{2}(\partial_{\mu}\phi)^2 + 
\frac{1}{2}M_{\phi}^2\phi^2 - \frac{\lambda}{4}\phi^4 +
\frac{1}{2}(\partial_{\mu}\chi)^2 + 
\frac{1}{2}M_{\chi}^2\chi^2 
- \frac{3\,g}{2}\phi^2\,\chi^2,
\end{eqnarray*}
where $\lambda$ and $g$ are coupling contants. In this work we study the influence of boundary conditions on energy levels of the field $\chi$ taking into account its interaction with field $\phi$. From above la-

\vspace*{0.8cm}
\noindent
1) {\it Instituto de F\'{\i}sica ``Gleb Wathagin'', University of Campinas (UNICAMP) - 13.081-970 - Campinas (SP), Brazil.}\\          
2) {\it Instituto de Matem\'atica, University of Campinas (UNICAMP) - 13.081-970 - Campinas (SP), Brazil; Ph: (55)-19-7885950; email: maia@ime.unicamp.br.}\\

\noindent
grangian we derive the following equations of motion for both fields.
\begin{eqnarray}
-\partial^{\mu}\partial_{\mu}\chi + M_{\chi}^{2}\chi - 3\,g\phi^{2}\chi = 0,
\label{sn1}
\end{eqnarray}
\begin{eqnarray}
-\partial^{\mu}\partial_{\mu}\phi + M_{\phi}^{2}\phi - \lambda\phi^{3}= 0.
\label{sn2}
\end{eqnarray}

In Eq.(2) we have neglected the term $3\,g\phi\chi^2$ which can be interpreted as the back-reaction of field $\chi$ on the mass term of $\phi$. This can be achieved if we impose that 
$|\chi|<< min\{\frac{M_{\phi}}{\sqrt{3\,g}}\}$.
Of course other regimes can be studied from Eq. (\ref{sn2}) adopting different approximations. 

It is well known the static solutions of the classical equation of motion (\ref{sn2}) are given by (Dashen {\it et al}, 1974)
\begin{eqnarray*}
\phi(x) = \pm\frac{M_{\phi}}{\sqrt{\lambda}}tanh(\frac{M_{\phi}x}
{\sqrt{2}}).
\end{eqnarray*}

These are Kink solutions which connect two vacua at $x=\pm\infty$. In this work unlike $\chi$, the field $\phi$ is not changed by boundary conditions. So in our approximation they are transparent to $\phi$. In a future work we will consider the most general case.

Substituting these solutions in Eq.(\ref{sn1}) we obtain 
\begin{eqnarray*}
-\partial^{\mu}\partial_{\mu}\chi + M_{\chi}^{2}\chi - 
3\frac{g}{\lambda}M_{\phi}^{2}tanh^{2}(\frac{M_{\phi}x}{\sqrt{2}}) \chi = 0.
\end{eqnarray*}

Since we are interested in stationary solutions, we can write $\chi(x) = 
e^{-i\omega\,t}\psi(x)$, where $\omega$ are energy eigenvalues and the 
previous equation reduces to
\begin{eqnarray}
\frac{d^2}{dx^{2}}\psi(x) + \left( M_{\chi}^2 + \omega^2 - 
3\frac{g}{\lambda}M_{\phi}^2
\,tanh^2(\frac{M_{\phi}x}{\sqrt{2}})\right)\psi(x) = 0.
\label{sn3}
\end{eqnarray}

This equation is similar to that one from DHN model (Dashen {\it et al}, 1974), which describe Kinks in $(1 + 1)$ dimensions, but here two different mass parameters appear in the potential.

In the section 2, we calculate the bounded energy levels of the field $\chi$ (or $\psi$) constrained to a box in (1 + 1) dimension with Dirichlet boundary conditions. We discuss how we can interpret the splitting of the ground state when the
box is shortened below a critical value, as particle production from a vacuum condensate. 

\vspace*{0.8cm}

\noindent
{\large \bf 2. BOUNDED SPECTRUM}

\vspace*{0.5cm}

In this section we obtain the energy levels of field $\chi$ 
imposing Dirichlet boundary conditions. This is as follows.
With the changing of variables
$z = \frac{M_{\phi}x}{\sqrt{2}}$ \ and \ $\omega^2 = 
\frac{(\varepsilon - 2)M_{\phi}^{2}}{2}$ \  
the equation (\ref{sn3}) can be written as:
\begin{eqnarray*}
\frac{d^2}{d z^2}\psi(z) + \left( 2\frac{M^{2}_{\chi}}{M^{2}_{\phi}} + 
\varepsilon - 2 - 6\frac{g}{\lambda}tanh^{2}z\right)\psi(z) = 0.
\end{eqnarray*}

Making a new changing of variable, namely,
$E = 2\beta + \varepsilon - 2$, where we have defined the mass ratio
$\beta\equiv\frac{M^{2}_{\chi}}{M^{2}_{\phi}}$, the above equation
reduces to 
\begin{eqnarray}
\frac{d^2}{d z^2}\psi(z) + (E - 6\frac{g}{\lambda}tanh^{2}z)\psi(z) = 0.
\label{sn4}
\end{eqnarray}

For discrete (bounded) spectrum case, $0\leq\omega^2 < 2M_{\chi}^2$, we have that $2\beta \leq E < 6\beta$.

In order to find the corresponding solution of the equation (\ref{sn4}) we make the following variable dependent transformation (Morse and Fesbach, 1953):
\begin{eqnarray}
\psi(z) = sech^{k}(z)Y(z),
\label{sn5}
\end{eqnarray}
where the parameter $k\in\mathbf{R}$ will be determined below.

Substituting (\ref{sn5}) in (\ref{sn4}) we get an equation for $Y(z)$
\begin{eqnarray}
\frac{d^2 Y}{d z^2}- 2\,k\,tanh(z)\frac{dY}{dz}+\left[ \frac{k^2+E-
6\frac{g}{\lambda}}{sech^{2}(z)}
+6\frac{g}{\lambda}-k^2-k\right]sech^2(z)Y(z) = 0.
\label{sn6}
\end{eqnarray}

In this work, we discuss the particular case $g=\lambda$. The general case, including an asymptotic study for weak and strong coupling constant $g$ will be presented elsewhere. Nevertheless the 
above condition leads to interesting results.
Firstly it is possible to turn this equation into a hypergeometric differential equation. This is as follows:

1.- We impose the term dependent on the variable $z$ in square brackets, 
equal to zero, which gives a relation between $k$ and $E$, i.e.,
\begin{eqnarray}
k = \pm\sqrt{6 - E}.
\label{sn7}
\end{eqnarray}

2.- Making a changing of the independent variable, namely,
\begin{eqnarray}
\mu = \frac{1}{2}(1 - tanh(z))
\label{sn8}
\end{eqnarray}
we obtain a hypergeometrical differential equation
\begin{eqnarray}
\mu(1-\mu)\frac{d^2 Y}{d\mu^2}+(k+1-2(k+1)\mu)\frac{dY}{d\mu}-
(k+3)(k-2)Y(\mu) = 0.
\label{sn9}
\end{eqnarray}

This equation has regular singular points at $\mu =0$, $\mu =1$ and $\mu = \infty$, and parameters $a=k+3$, $b=k-2$ and $c=k+1$. The two independent analytic solutions are given by (Gradshteyn and Ryzhik, 1980): 
\begin{eqnarray}
Y_1(\mu) = \,_2F_1(k+3,k-2;k+1;\mu), 
\label{sn10}
\end{eqnarray}
\begin{eqnarray}
Y_2(\mu) = \mu^{-k}\,_2F_1(-2,3;1-k;\mu).
\label{sn11}
\end{eqnarray}

As it is well known, if the parameter $c$ is a positive integer a
solution of the equation (\ref{sn9}) will be $Y_1(\mu)$ and, if $c$ is a 
negative integer, it will be given by $Y_2(\mu)$. Also there are other 
solutions containing a logarithmic term (Fuchs Theorem). If $c$ is a 
non-integer number, the set of two solutions $Y_1(\mu)$ and $Y_2(\mu)$ is a system of linearly independent solutions (Butkov, 1968). Next we study these three different cases for $c$. From now on, in order to simplify the notation, we denote the hypergeometric function $_2F_1$ simply as $F$.

\vspace*{0.5cm}

{\bf CASE I: $c$ Is a Positive Integer.}

\vspace*{0.3cm}

In this case, $c=k+1=n$, where $n=1,2,3,\dots$, and by Fuchs Theorem
(Butkov, 1968), the general solution is given by
\begin{eqnarray*}
Y(\mu) = AY_1(\mu)+BY_1(\mu)\ln\mid\mu\mid + B\sum_{s=0}^{\infty}a_{s}(-k)
\mu^{s-k},
\end{eqnarray*}
where $A$ and $B$ are arbitrary constants.

It is easy to see that this solution has a logarithmic divergence at $\mu = 0$. Since we want the solution $Y(\mu)$ be bounded, we must take $B=0$. This comes from the constraint that for a large box ($L\rightarrow\infty$) we must recover DHN's Kink and its bounded energy levels. So the solution reduces to 
\begin{eqnarray*}
Y(\mu)=A\,F(n+2,n-3;n;\mu),
\end{eqnarray*}
where $k$ was substituted by $n-1$.

Using the relations (\ref{sn5}) and (\ref{sn7}), as well changing
of variable $z=\frac{M_{\phi}x}{\sqrt{2}}$ in the above expression, we obtain
\begin{eqnarray}
\psi(x) = A\,sech^{(n-1)}(\frac{M_{\phi}x}{\sqrt{2}})
\,F(n+2,n-3;n;\frac{1}{2}\{1-tanh(\frac{M_{\phi}x}{\sqrt{2}})\}).
\label{sn12}
\end{eqnarray}

Below, in order to simplify the notation, we will denote $M_{\phi}$ as $M$.

Now we impose Dirichlet boundary conditions at $x=\pm\frac{L}{2}$ for the solution (\ref{sn12}), that is
\begin{eqnarray*}
\psi(\mp\frac{L}{2}) = A\,sech^{(n-1)}(\frac{ML}{2\sqrt{2}})
\,F(n+2,n-3;n;\frac{1}{2}\{1\pm tanh(\frac{ML}{2\sqrt{2}})\}) = 0.
\end{eqnarray*}

From these relations we get the condition 
\begin{eqnarray}
F(n+2,n-3;n;\frac{1}{2}\{1\pm tanh(\frac{ML}{2\sqrt{2}})\}) = 0.
\label{sn13}
\end{eqnarray}

On the other hand, since $2\beta \leq E < 6\beta$, from equation (\ref{sn7}), we can show that the parameter $k$ satisfies the inequalities
\begin{eqnarray*}
\sqrt{6(1-\beta)}<k\leq \sqrt{6 - 2\beta} \ \ \ \ \ \ \ \ \mbox{or} \ \ \ \ \ \
\ \
-\sqrt{6 - 2\beta} \leq k < -\sqrt{6(1-\beta)}.
\end{eqnarray*}

Since  we have considered that $k\in\mathbf{R}$, then from the above relations we obtain $\beta\leq 1$.

So, the possible values of parameter $k$ are in the intervals $0<k\leq \sqrt{6}\sim 2,44$ \ or \ $-2,44 \sim -\sqrt{6}\leq k<0$.
As in this case $c$ is a positive integer, we must take the interval $0<k \leq  \sqrt{6}$.

The allowed integer values of $k$ in the interval $0<k \leq  \sqrt{6}$ and the correspondent values of $E$ (from Eq.(\ref{sn13})) are (using the relation $k=n-1$)
\begin{eqnarray*}
\rm{for} \ \ \ \ n = 2, \ \ \ \ k = 1 \ \ \ \ \mbox{then} \ \ \ \ E = 5, 
\end{eqnarray*}
\begin{eqnarray}
\rm{for} \ \ \ \ n = 3, \ \ \ \ k = 2 \ \ \ \ \mbox{then} \ \ \ \ E = 2.  
\label{sn14}
\end{eqnarray}

The hypergeometric function in (\ref{sn13}) can be written as Jacobi Polinomials (Abramowitz and Stegun, 1972) and it is not difficult to prove that substituting the allowed values of $k$ (\ref{sn14})  into (\ref{sn13}) we do not get any consistent solution.

\vspace*{0.5cm}

{\bf CASE II: $c$ Is a Negative Integer.}

\vspace*{0.3cm}

Now we consider $c=k+1=-n$, where $n=1,2,3,\dots$, and then we must consider the interval $-\sqrt{6}\leq k<0$. The solution of the hypergeometric equation, in this case, is given by $Y_2(\mu)$. As for the previous case, by Fuchs Theorem, we have that the general solution is given by
\begin{eqnarray*}
Y(\mu) = A\,Y_2(\mu)+B\,Y_2(\mu)\ln\mid\mu\mid + B\sum_{s=0}^{\infty}a_{s}(-k)
\mu^{s-k},
\end{eqnarray*}
where $A$ and $B$ are arbitrary constants.

In this case we can not discard the $B$-terms in the same way as in the previous case, since for $\mu\rightarrow 0$ we have $\mu^{-k}\ln\mu\rightarrow 0$ and
$s-k>0$ for $s=0,1,2,\cdots .$ So we don't have a singularity at 
$\mu = 0$. Nevertheless, we notice that the relation 
$\frac{Y(\mu)}{Y_{2}(\mu)}$ is divergent in the asymptotic limit of 
$\mu = 0$, i.e.
\begin{eqnarray*}
\lim_{\mu\rightarrow 0}\frac{Y(\mu)}{Y_{2}(\mu)} \rightarrow \infty.
\end{eqnarray*}

Again, we can impose our natural boundary condition, that is, for very large box ($L\rightarrow\infty$), the DHN solution (Dashen {\it et al}, 1974) should be recovered.  In order to do this, we must impose 
the following asymptotic condition
\begin{eqnarray*}
\lim_{\mu\rightarrow 0}\frac{Y(\mu)}{Y_{2}(\mu)} \rightarrow 1.
\end{eqnarray*}

In order to this condition be valid, the  coefficient $B$ must vanish. 
Therefore, our solution reduces to
\begin{eqnarray*}
Y(\mu) = A\,\mu^{(n+1)}\,F(-2,3;2+n;\mu),
\end{eqnarray*}
where $k$ was substituted by $-(n+1)$, and $A\neq 0$.

As we did before, using the relations (\ref{sn5}) and (\ref{sn7}), as well the changing of variable $z=\frac{Mx}{\sqrt{2}}$ in the above expression, we can write the solution in the original variables:
\begin{eqnarray}
\psi(x) = A\,sech^{-(n+1)}(\frac{Mx}{\sqrt{2}})
\frac{\{1-tanh(\frac{Mx}{\sqrt{2}})\}^{(n+1)}}{2^{(n+1)}}
\,F(-2,3;2+n;\frac{1}{2}\{1-tanh(\frac{Mx}{\sqrt{2}})\}).
\label{sn15}
\end{eqnarray}

Now, imposing Dirichlet boundary conditions at $x=\pm\frac{L}{2}$, namely
\begin{eqnarray*}
\psi(\mp\frac{L}{2}) = A\,sech^{-(n+1)}(\frac{ML}{2\sqrt{2}})
\frac{\{1\pm tanh(\frac{ML}{\sqrt{2}})\}^{(n+1)}}{2^{(n+1)}}
\,F(-2,3;2+n;\frac{1}{2}\{1\pm tanh(\frac{ML}{2\sqrt{2}})\}) = 0,
\end{eqnarray*}
we obtain the condition
\begin{eqnarray}
F(-2,3;2+n;\frac{1}{2}\{1\pm tanh(\frac{ML}{2\sqrt{2}})\}) = 0.
\label{sn16}
\end{eqnarray}

In this case, the allowed integer values of $k$, in the interval $-\sqrt{6} \leq k < 0$ are using the relation $k=-(n+1)$ given by
\begin{eqnarray*}
\rm{for} \ \ \ \ n = 0, \ \ \ \ k = -1 \ \ \ \ \ \ \mbox{then} \ \ \ \ \ \ E = 5, 
\end{eqnarray*}
\begin{eqnarray}
\rm{for} \ \ \ \ n = 1, \ \ \ \ k = -2 \ \ \ \ \ \ \mbox{then} \ \ \ \ \ \ E = 2.  
\label{sn17}
\end{eqnarray}

As in the previous case it is not difficult to write the above hypergeometric function (\ref{sn16}) as Jacobi Polinomials (Abramowitz and Stegun, 1972). A quick analysis shows that no consistent solution exists for finite $L\neq 0$. For $L=\infty$, the DHN's case ($n=0$) is obtained.

\vspace*{0.5cm}

{\bf CASE III: $c$ Is a Non Integer.}

\vspace*{0.3cm}

Since $c$ is a non-integer number (positive or negative), the general solution is given by
\begin{eqnarray}
Y(\mu) = A\,F(k+3,k-2;k+1;\mu) + B\,(\mu)^{-k}\,F(-2,3;1-k;\mu).
\label{sn18}
\end{eqnarray}

Now we repeat the previous steps, i.e., using the relations (\ref{sn5}) and (\ref{sn7}), as well  the change of variable $z=\frac{Mx}{\sqrt{2}}$ in the previous expression, and obtain
\begin{eqnarray*}
\psi(x)  = sech^{k}(\frac{Mx}{\sqrt{2}})\left( A\,
F(k+3,k-2;k+1;\frac{1}{2}\{1-tanh(\frac{Mx}{\sqrt{2}})\}) \ + \right. 
\end{eqnarray*}
\begin{eqnarray}
\left. B\,2^{k}[1-tanh(\frac{Mx}{\sqrt{2}})]^{-k}\,F(-2,3;1-k;
\frac{1}{2}\{1-tanh(\frac{Mx}{\sqrt{2}})\})\right).
\label{sn19}
\end{eqnarray}

Now must determine $k$ and, afterwards, using relation (\ref{sn7}),
find the allowed values of $E$. The Dirichlet boundary conditions at $x=\pm\frac{L}{2}$ are given by
\begin{eqnarray*}
\psi(\mp\frac{L}{2})  =  sech^{k}(\frac{ML}{2\sqrt{2}})\left( A
\,F(k+3,k-2;k+1;\frac{1}{2}\{1\pm tanh(\frac{ML}{2\sqrt{2}})\}) \ + \right. 
\end{eqnarray*}
\begin{eqnarray*}
\left. B\,2^{k}[1\pm tanh(\frac{ML}{2\sqrt{2}})]^{-k}\,F(-2,3;1-k;
\frac{1}{2}\{1\pm tanh(\frac{ML}{2\sqrt{2}})\})\right) = 0.
\end{eqnarray*}

From these relations we obtain 
\begin{eqnarray*}
A\,F(k+3,k-2;k+1;\frac{1}{2}\{1\pm tanh(\frac{ML}{2\sqrt{2}})\}) \ + 
\end{eqnarray*}
\begin{eqnarray*}
B\,2^{k}[1\pm tanh(\frac{ML}{2\sqrt{2}})]^{-k}\,F(-2,3;1-k;
\frac{1}{2}\{1\pm tanh(\frac{ML}{2\sqrt{2}})\}) = 0.
\end{eqnarray*}

This is a system of homogeneous equations for $A$ and $B$. So this system has a non-trivial solution only if the determinant of system is zero, i.e.,
\begin{eqnarray*}
\{1-tanh(\frac{ML}{2\sqrt{2}})\}^{-k}
\,F(k+3,k-2;k+1;\frac{1}{2}\{1+tanh(\frac{ML}{2\sqrt{2}})\})\times  
\end{eqnarray*}
\begin{eqnarray*}
\,F(-2,3;1-k;\frac{1}{2}\{1-tanh(\frac{ML}{2\sqrt{2}})\}) \ -
\{1+tanh(\frac{ML}{2\sqrt{2}})\}^{-k} \times 
\end{eqnarray*}
\begin{eqnarray}
\,F(k+3,k-2;k+1;\frac{1}{2}\{1-tanh(\frac{ML}{2\sqrt{2}})\})
\,F(-2,3;1-k;\frac{1}{2}\{1+tanh(\frac{ML}{2\sqrt{2}})\}) = 0.
\label{sn20}
\end{eqnarray}

Again we can write the above hypergeometric functions as Jacobi Polinomials (Abramowitz and Stegun, 1972). In this way, from (\ref{sn20}) we obtain a transcendental equation for the parameter $k$, i.e.,
\begin{eqnarray}
{\left(\frac{1+tanh(\frac{ML}{2\sqrt{2}})}{1-tanh(\frac{ML}{2\sqrt{2}})}\right)}
^{k} =\pm\left(\frac{k^2 - 1 + 3\,k\,tanh(\frac{ML}{2\sqrt{2}}) +
3\,{tanh}^{2}(\frac{ML}{2\sqrt{2}})}{k^2 - 1 - 3\,k\,tanh(\frac{ML}{2\sqrt{2}})
 + 3\,{tanh}^{2}(\frac{ML}{2\sqrt{2}})}\right).
\label{sn21}
\end{eqnarray}

Notice that substituting $k$ by $-k$ in equation (\ref{sn21}), it can be 
verified that this equation is satisfied. So these solutions are valid in the intervals $(0,1)\cup(1,2)\cup(2,\sqrt{6})$ $\subset \mathbf{R}$ \ and \ $(-\sqrt{6},-2)\cup(-2,-1)\cup(-1,0)$ $\subset \mathbf{R}$. Observe that for the negative sign, $k = -1$ is a solution of equation (\ref{sn21}), but it is not allowed since now we are considering $k\not\in\mathbf{Z}$. 

An analytic solution for the transcendental equation (\ref{sn21}) was not founded but it is possible to obtain numerical solutions for it as
we can see in the figure below for the case that $\beta =1$.

\vspace*{0.8cm}

\begin{figure}[ht]
\centerline{\psfig{figure=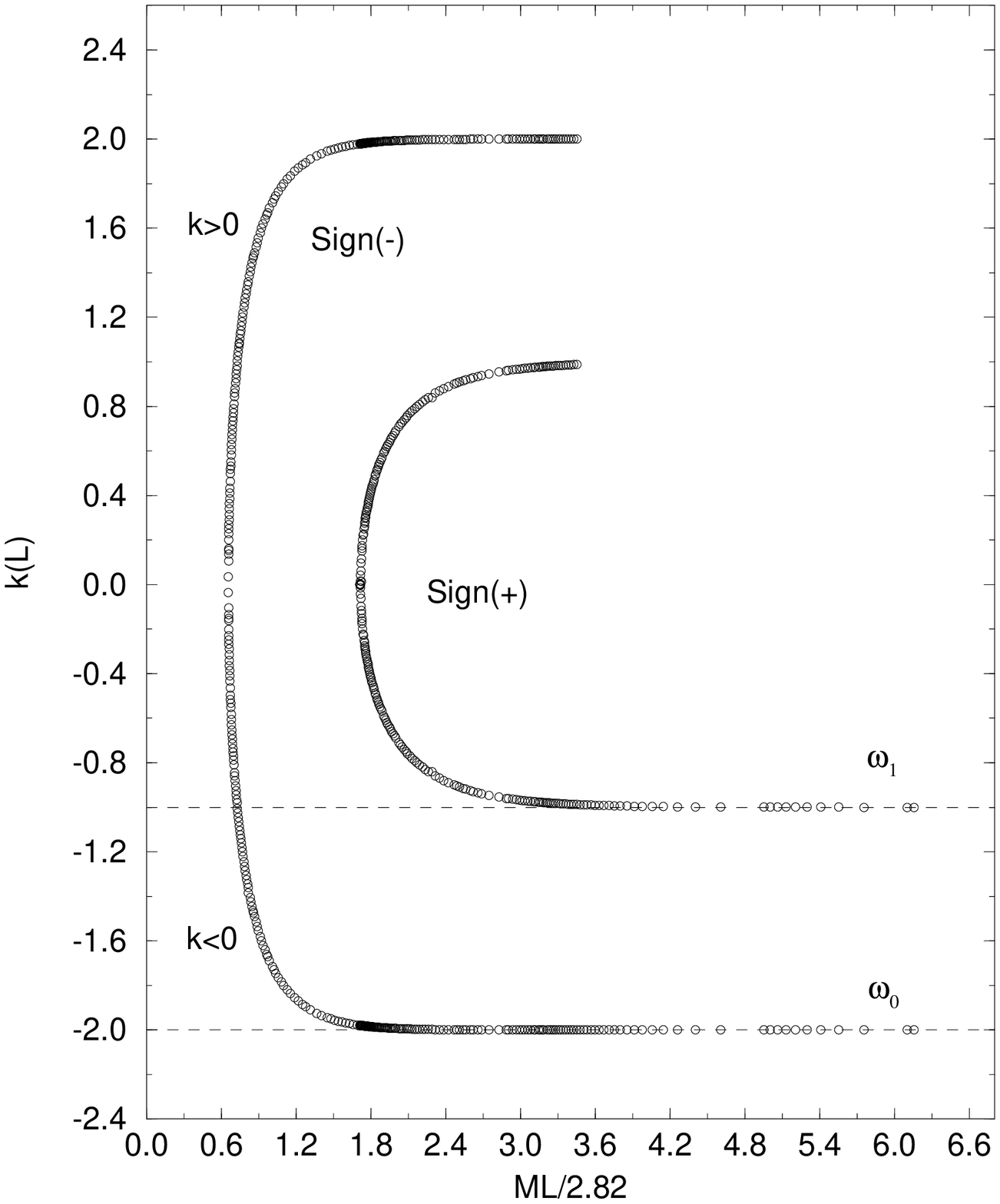,height=3.5in,width=3in}}
\end{figure}
\begin{center}
\noindent
\parbox{13cm}{{\bf Fig. 1.} Shifts of Bound States with the size of the Box. Here $\beta =1$. The horizontal lines show the asymptotic values of the DHN model (Dashen {\it et al}, 1974).}
\end{center}

\vspace*{0.5cm}
 
The Fig. $1$ shows the relation between parameter $k$ and the size $L$ of the box. Observe that both ground state and the first excited state shift with the size of the box. Starting with a very large box size $L$, as it decreases, the values of $\omega_0$ and $\omega_1$ increases from 
$\omega_0 = 0$ $(k=-2 \ \mbox{and} \ L\rightarrow\infty)$ to $\omega_0 = \sqrt{2}M$ $(k=0)$, and from $\omega_1 =\sqrt{\frac{3}{2}}M$ $(k=-1, \ \mbox{and} \ L\rightarrow\infty)$ to $\omega_1 =\sqrt{2}M$ $(k=0)$. Close to critical value $\sim 0.6$ for $\omega_0$, and $\sim 1.71$ for $\omega_1$ these bound states merge in the continuum part of the spectrum $(k=0)$. On the other hand, for $L\rightarrow\infty$ the values of $\omega_0$ and $\omega_1$ decreases until to reach their minimum values $\omega_0 =0$ and $\omega_1 =\sqrt{\frac{3}{2}}M$ respectively, just as in the DHN model. This behaviour happens for both cases of positive and negative signs in the equation (\ref{sn21}) with negative $k$. For any positive $k$ and large box sizes $L$ the equation (\ref{sn21}) has no solution, which can be checked analitically 
(see also the Fig.1 above). So we discard positive values of $k$ since for a large box ($L\rightarrow\infty$) they do not
lead to DHN's energy levels.

We can also obtain a relation between $\frac{\omega(L)}{M}$ and
box size $L$. From equation (\ref{sn7}) and the relation
$\omega^2 =\frac{(\varepsilon -2)M^2}{2}$ we get
\begin{eqnarray}
\frac{\omega}{M} = \pm\sqrt{3 - \beta -
\frac{k^2(L)}{2}}.
\label{sn22}
\end{eqnarray}

In the Fig. $2$ below we plotted two cases for this relation, namely
$\beta=1$ and \ $\beta =\frac{1}{2}$ .

\vspace*{0.8cm}

\begin{figure}[ht]
\centerline{\psfig{figure=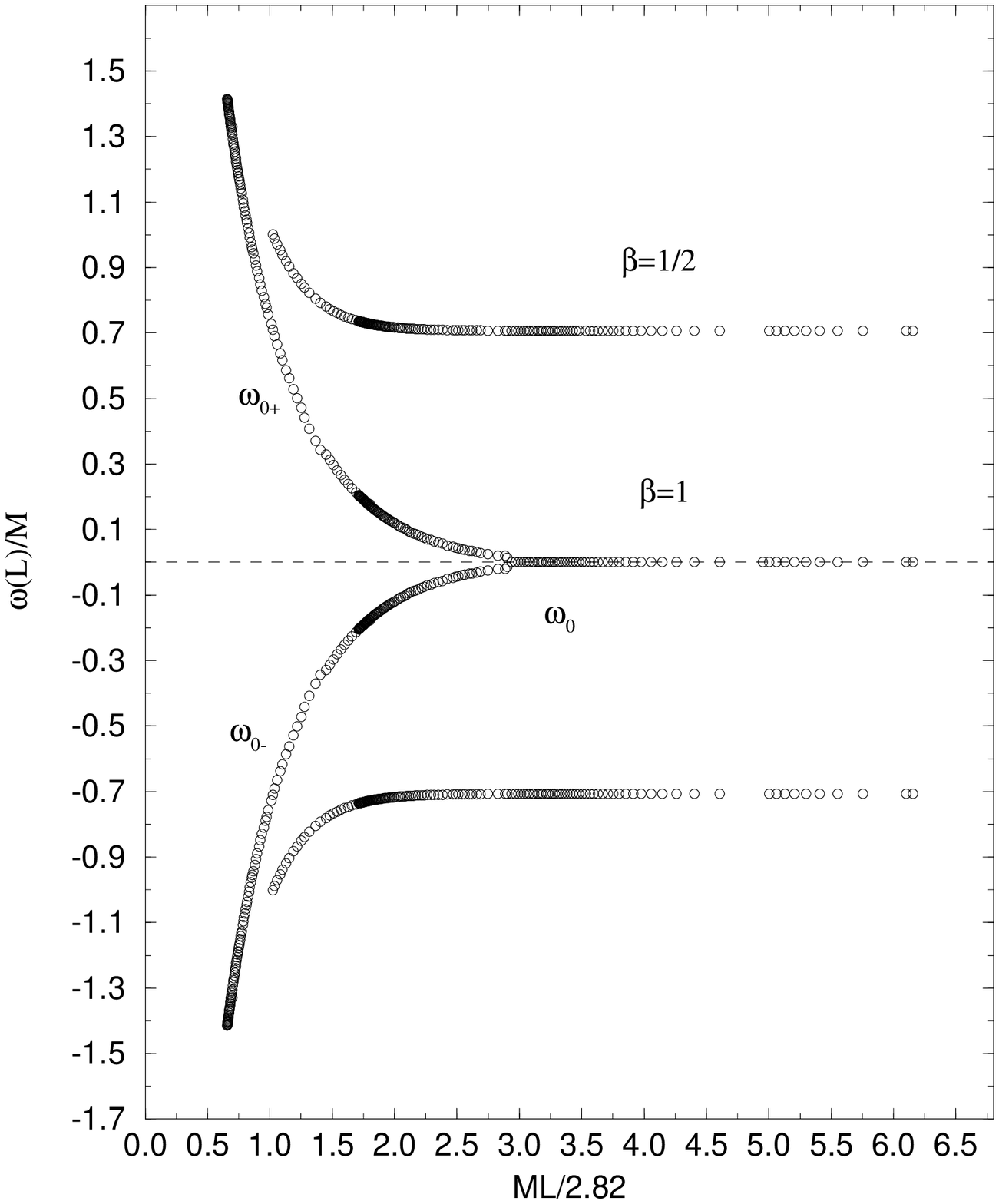,height=3.5in,width=3in}}
\end{figure}
\begin{center}
\noindent
\parbox{13cm}{{\bf Fig. 2.} Energy levels of the ground state of field $\chi$ 
(or $\psi$) dependent of the box size $L$. The horizontal lines show the asymptotic values of DHN model (Dashen {\it et al}, 1974).}
\end{center}

\vspace*{0.5cm}

Now we can see that for the first case $(\beta=1)$, the ground state $\omega_0$, for large box size may form an aproximated vacuum particle-antiparticle condensate, since both levels are exponentially close one from another. Roughly speaking below a critical size ($\sim 2.93$) any perturbation of box size will induce pair formation from this energy level. This could be important for particle production in the presence of strong fields. Squeezed fields could have an enhanced production.

Another interesting behaviour of the energy levels is showed in Fig. $3$ below. We plotted the difference between the levels $\omega_{0}$ and $\omega_{1}$, for arbitrary distance $L$. The interesting fact about this is the peak around the critical value $\frac{ML}{2\sqrt{2}}\sim 2.93$ and that the increasing part of the curve (left-hand side) from the peak shows a not smooth, growth with several secondary maxima and minima.

\newpage

\begin{figure}[ht]
\centerline{\psfig{figure=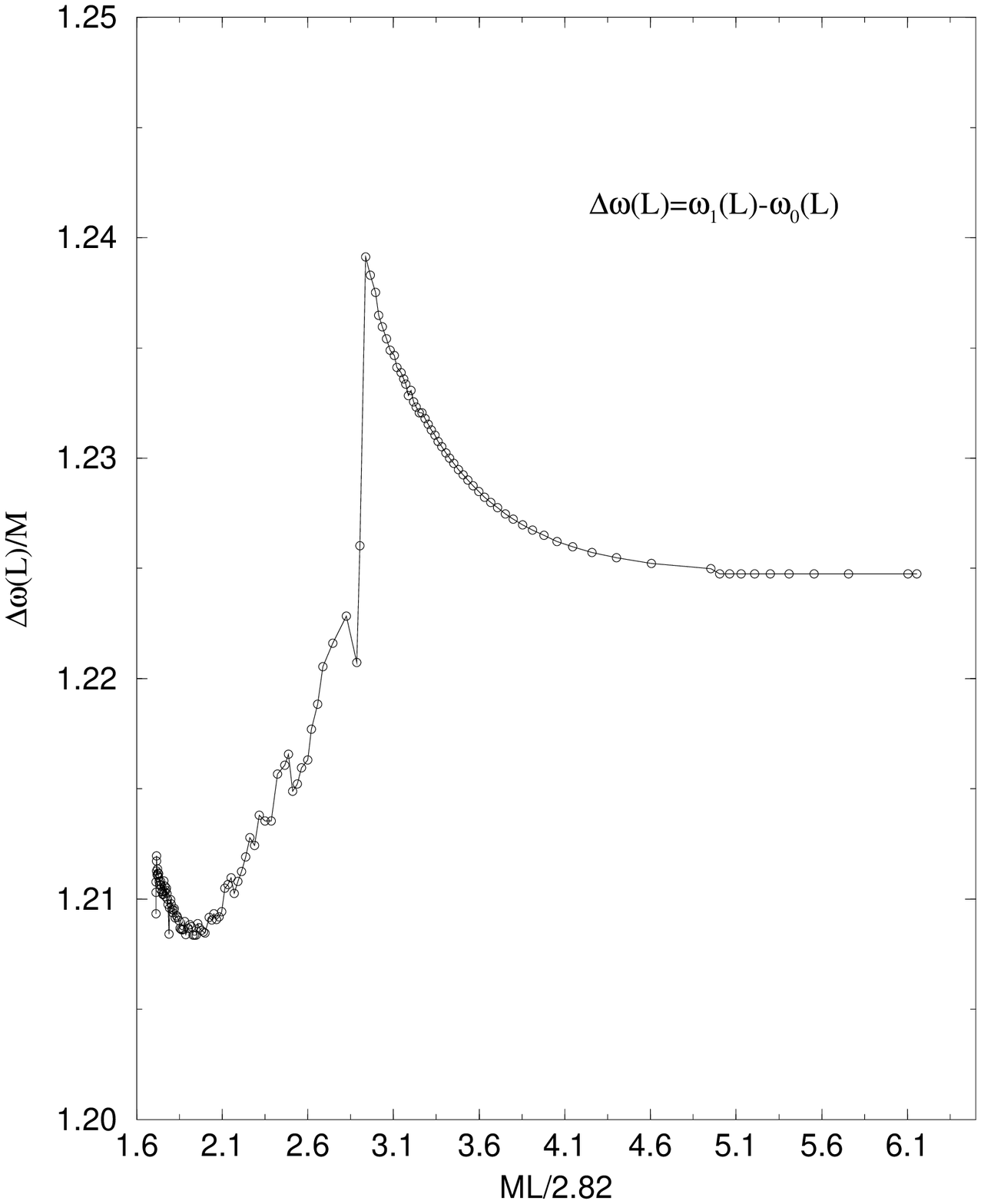,height=3.5in,width=3in}}
\end{figure}
\begin{center}
\noindent
\parbox{13cm}{{\bf Fig. 3.} The gap between levels $\omega_0$ and $\omega_1$ versus the size $L$ of the box.}
\end{center}

\vspace*{0.8cm}

\noindent
{\large \bf 3. CONCLUSIONS}

\vspace*{0.5cm}

In this work we calculated the solutions of a Klein-Gordon type equation in an one-dimensional 
box for which we impose Dirichlet boundary conditions, and in the presence of a Kink-type potential generated by a second scalar, self-interacting field which, in our approximation, is not subjected to the boundary conditions. Energy levels of bounded states for non-trivial solutions are obtained as roots of a transcendental equation involving $L$ and $k$. Although we have not obtained analytic solutions to it, we have studied numerical solutions (see Fig. $1$). 

The ground state $\omega_0$ and the first excited state $\omega_1$ shifts with the size $L$ of the box (see fig. 1), for both cases of positive and negative signs. As the size $L$ of the box decreases, $\omega_0$ increases in the interval 
$[0, \sqrt{2}M]$, and $\omega_1$ in $[\sqrt{\frac{3}{2}}M, \sqrt{2}M]$. Close to the critical value $\sim 0.6$ for $\omega_0$, and $\sim 1.71$ for $\omega_1$, all the bounded states merge in the continuum part of the spectrum. For large distances ($L=\infty$) we obtain the energy levels of DHN model. The decrease of $L$ induces shifts on the bound states levels of the system, and close to a critical size $\sim 2.93$ we have ``just barely bound'' condensate (Morse and Feshbach, 1953) that may decay against a small perturbation on the system, with particle pair creation (Fig. 2). 

The gap between the two bound states presents a peak at $\frac{ML}{2\sqrt{2}}\sim 2.93$ and shows a wildly (non-smooth) behaviour. It is interesting that the critical value for the splitting of the levels coincide with the value to the peak position, but we do not have any explanation for this fact up to now. This, as well the study of the system taking into account finite boundary conditions to both fields $\phi$ and $\chi$ will be done elsewhere.

\vspace*{0.8cm}

\noindent
{\large \bf ACKNOWLEDGMENTS}

\vspace*{0.5cm}

This work was supported, in part, by  FAPESP (Funda\c{c}\~ao de 
Amparo \`a Pesquisa do Estado de S\~ao Paulo), Brazil. The authors are indebted to Professor A. A. Grib for his 
kind suggestions and discussions.

\vspace*{0.8cm}

\noindent
{\large \bf REFERENCES}

\vspace*{0.5cm}
\noindent
Abramowitz M., and Stegun I. A. (1972). {\em Handbook of Mathematical Functions}, Dover Publications, INC, New York. \\
Butkov E. (1968). {\em Mathematical Physics}, Addison-Wesley
Publishing Company. \\
Dashen R., Hasslacher B., and Neveu A. (1974). {\em Phys. 
Rev. {\bf D10}}, 4131. \\
Gradshteyn I.S., and Ryzhik I.H. (1980). {\em Table of Integrals, Series and Products}, Academic Press, New York. \\
Morse P.M., and Feshbach H. (1953). {\em Methods of Theoretical
Physics}, McGraw-Hill, New York.

\end{document}